# The origins of length contraction: I. The FitzGerald-Lorentz deformation hypothesis

Harvey R Brown
*Sub-Faculty of Philosophy, University of Oxford, 10 Merton Street, Oxford OX1 4JJ, U.K.*

> "Can there be some point in the theory of Mr. Michelson's experiment which has yet been overlooked?" H.A. Lorentz, letter to Lord Rayleigh, August 1892.

One of the widespread confusions concerning the history of the 1887 Michelson-Morley experiment has to do with the initial explanation of this celebrated null result due independently to FitzGerald and Lorentz. In neither case was a strict, longitudinal length contraction hypothesis invoked, as is commonly supposed. Lorentz postulated, particularly in 1895, any one of a certain family of possible deformation effects for rigid bodies in motion, including purely transverse alteration, and expansion as well as contraction; FitzGerald may well have had the same family in mind. A careful analysis of the Michelson-Morley experiment (which reveals a number of serious inadequacies in many text-book treatments) indeed shows that strict contraction is not required.

1. Introduction

It is well-known that in providing a derivation of the inertial coordinate transformations based on the relativity principle and his light postulate, Albert Einstein obtained in 1905 a not unfamiliar result in a novel way. The same transformations had already effectively appeared in the context of the electron theory in the 1900 work of Joseph Larmor, and the independent 1904 work of H.A. Lorentz. Einstein showed in his celebrated 1905 relativity paper[1] that what Henri Poincaré had dubbed the "Lorentz transformations" lead to a longitudinal contraction for rigid bodies in motion. It is still widely believed that such an effect—if not its interpretation—corresponds precisely to that involved in the FitzGerald-Lorentz 'contraction' hypothesis which had appeared over a decade earlier.

But this assumption is not strictly correct. The discrepancy between Einstein's relativistic contraction prediction and the original Lorentz hypothesis has already been noted, as we shall see, in the work of a handful of commentators; in the case of FitzGerald the discrepancy is probably the same but it is almost never acknowledged in modern treatments. It is our intention here to spell out the nature and significance of this discrepancy in more detail than has hitherto been done. The motivation for this study is the belief that the transition from the original FitzGerald-Lorentz hypothesis to the correct form of the contraction prediction in 1905 (to be discussed in a subsequent paper) was an important and under-rated development in our understanding of the nature of kinematics.

The FitzGerald-Lorentz (FL) hypothesis was of course the result of a somewhat desperate attempt to reconcile the null result of the 1887 Michelson-Morley (MM) experiment with the hitherto successful Fresnel-Lorentz theory of a stationary luminiferous ether, a medium through which the earth is assumed to move with unappreciable drag. The MM experiment is rightly regarded today as one of the turning points in physics, and although it is discussed widely in textbooks, it is remarkable how much confusion still surrounds its structure and meaning. In order then to understand the FL hypothesis, it is necessary first to go over some well-trodden ground; sections 2 and 3 below are designed to show what the 1887 null result does and does not imply. In particular it is shown in section 3 that in the context of a theory of light in which the light-speed is independent of the speed of the source, a certain motion-induced deformation of rigid bodies, of which contraction is a special case, is required. (A common misunderstanding about the role of the Galilean law of addition, or transformation of velocities, in the analysis of the MM experiment is also exposed.) In section 4, this result is incorporated into a more systematic account of what we call MM kinematics, which is found to be consistent with the 1949 review by Robertson of the experimental underpinnings of relativistic kinematics.

In sections 5 and 6, the history of the independent responses of both FitzGerald and Lorentz to the MM experiment is reviewed, and it is argued that in neither case was a purely (longitudinal) contraction effect being proposed originally. Nor was the FL deformation hypothesis as artificial or *ad hoc* as it was (and is) often portrayed. In both cases, a plausible dynamical underpinning of the hypothesis was proposed, as is seen in sections 7 and 8. But the subtle, yet important difference between these dynamical arguments, each relying on an analogy with the effect of motion on electrostatic forces, is discussed in a subsequent paper, as is the history of the eventual shift from FL deformation to relativistic contraction.

2. The non-relativistic analysis of the Michelson-Morley experiment.

The keystone assumption in the ether theory of light was that relative to the inertial rest-frame of the ether, the 'two-way' (or 'round-trip') light-speed *in vacuo* is a 'constant' *c*, that is to say, *independent of the motion of the source of light and isotropic*. It was originally expected, on the basis of Galilean kinematics, that the two-way light-speeds (i.e both ray and phase speeds[2]) relative to an inertial frame moving with respect to the ether—such as that defined by the earth over sufficiently short periods of time—would likewise be source-speed independent *but not isotropic*. The 1887 experiment of Michelson and Morley was designed to detect the predicted anisotropy of the ray speed— the first attempt, following Michelson's similar but inconclusive effort in 1881, to detect a second-order optical effect of the ether wind.[3] (The null results in all previous first-order ether wind experiments were explained by appeal to the Fresnel drag coefficient for light-speeds inside moving transparent media, which H. A. Lorentz showed to be a consequence of his dynamical theory of the stationary ether.[4]) Famously, the shift in the interference pattern that was expected to appear in the process of rotating the Michelson optical interferometer through 90° was not observed.

Let *S* represent the inertial rest frame of the ether, and *S′* the inertial rest frame of the earth-bound laboratory. Relative to *S* the two-way light-speed is assumed to be isotropic, as we have seen. Thus we may naturally adopt for *S* the Einstein convention for synchronising distant clocks, which has the effect of rendering isotropic the *one-way* speed of light in that frame. For the moment we shall leave unspecified the corresponding convention in *S′*—nothing in this and the next section depends on it.

Initially at rest relative to *S′* is the MM interferometer, one of whose two, perpendicular arms (call it *A*) points in the direction of the motion of *S′* relative to *S*. As usual, we shall suppose that the coordinate systems adapted to *S* and *S′* are in the 'standard configuration', with the arm *A* pointing in direction of the the positive *x*-, *x′*-axes. The second arm *B* lies parallel to the *y′*-axis. With a view to later discussion, we analyse the general case in which the rest lengths of the two arms, $L_A′$ and $L_B′$, measured of course relative to the laboratory frame *S′*, are not equal.

Standard non-relativistic treatments of the experiment in the literature divide into two kinds: those which (like the 1887 MM account) calculate delay times etc. relative, in the first instance, to the ether rest frame *S*, and those which, by appeal to the Galilean rule for transformation of (ray) velocities, deal from the outset with such time intervals defined relative to the lab frame *S′*. We shall for the most part rehearse the first kind, for reasons that become clear shortly.

Relative to *S*, the speed of light in both directions in *A* is *c*. This claim follows immediately from the above assumption of the 'constancy' of the two-way light-speed relative to *S*, and the further Fresnel-Lorentz assumption that the interferometer does not drag the ether with it while in motion. A simple calculation shows that the time taken for a pulse of light to leave the beam splitter (semi-transparent mirror) at the join of the two arms and return to it *via* arm *A* is, relative to *S*,

$$T_A = 2 \gamma^2 L_A / c \tag{1}$$

where $L_A$ is the length of arm *A* relative to *S* and $\gamma$ is the now familiar 'Lorentz' factor $(1 - v^2/c^2)^{-1/2}$.

In calculating the corresponding traverse time along arm *B*, it is normally assumed that the (ray) velocities in this case have components $u_x = v$, $u_y = \pm(c^2 - v^2)^{1/2}$. Let us call this assumption the "*B*-claim". It now follows that

$$T_B = 2 \gamma L_B / c \tag{2}$$

where $L_B$ is the length of *B* relative to *S*.[5] (It should be noticed that both $T_A$ and $T_B$ are by definition measured by *pairs* of synchronised, separated clocks at rest relative to *S*; whereas their primed counterparts are measured by a *single* clock at rest relative to the interferometer.)

Consider now the delay time based on these expressions:

$$\Delta := T_A - T_B = 2\gamma(\gamma L_A - L_B)/c. \tag{3}$$

We wish to compare this delay $\Delta$ with that occurring after the interferometer is rotated by 90°, so that arm *A* comes to rest pointing in the positive *y'* direction, *B* in the negative *x'* direction. Similar reasoning yields for the new delay time $\Delta^{rot}$ (assuming the isotropy of space in *S*)

$$\Delta^{rot} := T_A^{rot} - T_B^{rot} = 2\gamma(L_A^{rot} - \gamma L_B^{rot})/c, \tag{4}$$

where $L_A^{rot}$ is the length relative to *S* of the rotated arm *A*, etc.[6] (Such at least is the standard claim, but as we shall see, it seems that (3) and (4) can not both be exact expressions without a re-adjustment of the angle of the beam-splitter during rotation.)

Notice that so far, no assumptions inconsistent with relativity theory have appeared. This is in contrast with the second type of common analysis mentioned above, in which a derivation is given of the expressions for pre- and post-rotation delay times defined from the start relative to the lab frame. As the speed of light is no longer expected to be isotropic relative to this frame, the speeds in the *A* and *B* arms are now provided by way of the Galilean rule for transforming ray velocities. The expressions for the delay times are identical to (3), (4) but in the primed variables. We shall return to this type of analysis and its potential misconstrual in the next section.

The eye-piece, or telescope, attached to the interferometer received, throughout the rotation, a superposition of two monochromatic beams of light emerging from the beam-splitter that were identical up to a certain phase difference; the mirrors were arranged to produce straight fringes. Suppose $n$ and $n^{rot}$ are the (frame-independent) number of periods of the light waves associated with the time delay in traversing the two arms, before and after rotation respectively. According to Michelson and Morley, and many subsequent commentators, the phase shift due to 90° rotation is then (in units measured by the distance between the fringes)

$$\Delta n := n - n^{rot} = (\tau - \tau^{rot})(c/\lambda) \qquad (5)$$

where $\lambda$ is the known wavelength of the sodium light.

Now the lengths of the arms of the MM interferometer relative to the laboratory frame were arranged to be very nearly equal ($L_A' = L_B' = L'$) and, in accordance with Galilean kinematics, Michelson and Morley assumed the absence of any motion-induced shape deformation ($L_A = L_B = L_A^{rot} = L_B^{rot} = L'$). Under these conditions, one obtains from (3), (4) and (5) the expression for $\Delta n$ up to second order in $v/c$:

$$\Delta n \sim 2(L'/\lambda)(v^2/c^2). \qquad (6)$$

Assuming that the orbital velocity of the earth relative to the sun coincides with the velocity $v$ of the laboratory relative to the ether, then for $L' \sim 11$m (the length obtained in the MM interferometer by means of multiple reflections in each arm), the prediction was $\Delta n \sim 0.4$ fringe. Michelson and Morley reported that the "actual displacement was certainly less than the twentieth part of this, and probably less than the fortieth part." But before we analyse the implications of this 'null' result, we should finish the story about what ought to be the exact non-relativistic prediction for the experiment.

First, note that equation (5) seems to be mixing quantities that are defined relative to different frames. The wavelength of light is always defined relative to the rest frame of the actual equipment—specifically the detector—used by the observer, and should be denoted $\lambda$. Thus it is best to understand the quantity $(\tau - \tau^{rot})$ in (5) as the relative time delay defined also with respect to the lab frame $S$. Under the *non-relativistic* assumption of zero time dilation, it is of course determined by the difference between the delay times in (3) and (4). But we face further problems if we insist on treating (5) as exact: we are now left with the (phase) velocity $c$ defined relative to the wrong frame, and the rest wavelength of the sodium light is apparently being treated as isotropic relative to $S$. This last assumption is inconsistent with the fact that the rest frequency of light is frame-independent, combined with the fact that the phase velocity is not isotropic relative to the lab frame $S$ according to nonrelativistic kinematics.

Clearly, (5) is not exact. It would be, however, if the factor $c/\lambda$ were replaced by $\nu^o$, the known rest frequency of the sodium light. (This is the same as $c/\lambda$ —the anisotropy factors in the numerator and denominator being the same, and hence cancelling—where $c$ is the phase speed of light relative to $S'$.) But since it is effectively the measured value of $\nu^o$ that Michelson and Morley are providing in estimating the fringe shift, such qualms about (5) are essentially academic.

Second, Michelson and Morley obtained the standard non-relativistic prediction (6) despite the fact that their 1887 expression for $T_B$ was actually distinct from (2)—a fact first recognised in the literature apparently only in 1989![7] The reasoning behind their incorrect expression ($T_B = (2L_B/c)(1 + v^2/c^2)^{1/2}$ in our notation) is a matter of speculation[8]; what is significant for our purposes is that it coincides to second order with (2) and hence yields (6).

Third, we note that in a detailed and provocative analysis of the MM experiment published in 1994, M. Mamone Capria and F. Pambianco claimed that like its original MM counterpart, (2) above is also correct only to second order! This new analysis relied on a systematic application of Huygens' principle to the case of reflection of plane light waves off moving mirrors, and was based on the assumption that the beam splitter is inclined at *exactly* 45° to the direction of the incoming beam. It was demonstrated that $T_B$ is given by[9]

$$T_B = (L_B/c)\{1 - v/c + (1 - v/c)^{-1}\}, \qquad (7)$$

which to second order coincides with (2). The analysis showed moreover that the paths of the two rays from the beam splitter to the telescope do not coincide, and in particular do not have the same optical length. When this detail is taken into account, the new, exact total delay time is

$$= T_A - T_B + (Fv/c)(1 - {}^2), \qquad (8)$$

where $T_A$ is given by (1), $T_B$ is given by (7), and $F$ is the distance from the $x$-axis to the telescope[10]. The discrepancy between this delay time and the standard one given by (3) (where $T_B$ is given by (2)) is, happily, only in third and higher order terms.

Returning to the standard account based on (2) and (3), it was argued above that the effect of the 90° rotation of the interferometer is simply to exchange the role of arms *A* and *B* in calculating the delay times. Mamone Capria and Pambianco showed, again under the exact 45° tilt angle assumption, that such symmetry again only strictly holds to second order, and then only when the assumption $L_A = L_B = L_A^{rot} = L_B^{rot} = L'$ is made.[11] The upshot is that whereas the standard fringe shift (6) survives to second order in this new non-relativistic analysis, the exact treatment therein looks considerably more complicated than its standard counterpart.

Is it right? Yes, but not in the sense of impugning the exact validity of (2). This expression was based only on the *B*-claim above, which in turn is based on the assumption that the transverse ray in *B* passes after reflection back through the mid-point of the beam splitter. In practice, the beam splitter is adjusted so as to maximise the definition of the interference fringes in the telescope. And although occasionally it is stated[12] in instances of the standard non-relativistic analysis in the literature that the (frame-independent) angle of tilt of the beam splitter is 45°, given the ether wind supposedly blowing through the apparatus it should not be expected to be exactly so[13]. Mamone Capria and Pambianco in fact showed what the precise tilt angle must be ($\sin^{-1}(2 - v^2/c^2)^{-1/2}$) for the simple formula (2) to hold exactly, and they anticipated the objection that the beam splitter could be adjusted to achieve this angle. They claimed that in this case a further tiny adjustment is required for (4) to hold exactly, and not just to second order.[14] Indeed, it is not hard to show that according to the non-relativistic analysis, the 90° rotation does not exactly preserve the 'ideal' angle of tilt of the beam splitter.

## 3. Deformation—the general explanation of the null result.

Let us now return to our previous considerations for the case where $L_A'$ is not necessarily equal to $L_B'$, and where Galilean kinematics are not assumed to hold. We define the longitudinal and transverse 'deformation factors' relative to the ether frame *S*:

$$C_L = L_A/L_A^\circ = L_B^{rot}/L_B^\circ \qquad (9)$$
$$C_T = L_B/L_B^\circ = L_A^{rot}/L_A^\circ, \qquad (10)$$

where $L_{A(B)}^\circ$ is the length of arm *A(B)* relative to *S*, when the interferometer is hypothetically at rest relative to that frame. (These deformation factors are expected to depend on *v* and of course to tend to unity in the limit $v \to 0$.) It follows from the isotropy of space relative to *S* that such rest lengths do not depend on the orientation of the interferometer. If it is further assumed that any length alterations due to motion hold for *all* the rigid bodies (including rulers) in the laboratory as well as for stone slab on which the interferometer is mounted, then rest-lengths will be invariant, or frame-independent, i.e.:

$$L_A^\circ = L_A', \quad L_B^\circ = L_B'. \qquad (11)$$

In particular, it follows from the two assumptions above that $L_A'$ and $L_B'$ also do not depend on the orientation of the interferometer.

Now the *exact* absence of any fringe shift under rotation (when not assuming $L_A' = L_B' = L'$) would require in the standard analysis that the pre- and post-rotation delay times given in (3) and (4) are equal. It follows in this case, using (9) and (10), that

$$\boldsymbol{C}_T = \boldsymbol{C}_L. \qquad (12)$$

Even if it is accepted that (3) and (4) cannot both be exact, an explanation of the null MM result ensues if the dimensions relative to *S* of any rigid body undergo an anisotropic change as a result of motion, which is consistent up to second order with (12) and hence with:

$$\boldsymbol{C}_T = (1 + v^2/2c^2)\boldsymbol{C}_L. \qquad (13)$$

Note that if such shape deformation is to occur, then the angle of tilt of the beam splitter will no longer be frame independent. We have mentioned that the wave-theoretical analysis of Mamone Capria and Pambiano fixes the value of the tilt angle to be $\sin^{-1}(2 - v^2/c^2)^{-1/2}$ relative to the ether frame *S* in order that the transverse ray in *B* passes back through the mid-point of the beam splitter—in other words to ensure the exact validity of (2). Unsurprisingly, it is also the angle given by applying the anisotropic deformation (12) directly to the semi-transparent mirror itself, when the angle relative to the lab frame *S* is exactly 45°![15]

Several further points must be stressed at this stage. First, such a motion-induced distortion or deformation (12) is, of course, more general than the purely longitudinal contraction effect we are familiar with in special relativity theory (which follows from (12) when $\boldsymbol{C}_T = 1$). The deformation may actually involve expansion and/or contraction effects (i.e. it is not required that $\boldsymbol{C}_L, \boldsymbol{C}_T$ be less than 1), but it cannot vanish *in toto*. Second, without appeal to the relativity principle, there is no reason to expect that the deformation factors will be frame-independent (as will be become evident below). Third, the existence of such an anisotropic deformation defined relative to *S* does *not* invalidate the assumption implicitly made by Michelson and Morley that relative to the laboratory frame *S'* the lengths of the two interferometer arms were equal *before and after rotation*. Fourth, the explanation above of the MM null result involving rigid-body deformation does not depend on prior knowledge of the form of the coordinate transformations between the frames *S* and *S'*, nor even on the associated rule for transformation of ray velocities. However, the existence of such a universal deformation will of course act as a constraint on the form of the coordinate transformations, as is seen in the next section.

Very few textbook treatments of the MM experiment recognise anything other than a purely longitudinal contraction as an explanation of the null result (the 1994 treatment by Mills being a notable recent exception[16]). However, the failure in several treatments to recognise the third and fourth points above is arguably even more alarming. Mention was made in the previous section of a common textbook analysis of the MM experiment making use of the Galilean law for transformation of ray velocities (for both longitudinal and transverse components) in order to calculate directly the delay times $\delta'$, $(\delta^{rot})'$ as defined relative to the laboratory frame *S'*, assuming as always light-speed constancy in *S*. Given this law, in the case where the arms of the interferometer are of equal length the primed delay times are readily shown to be:

$$\delta' = -(\delta^{rot})' = 2(\gamma - 1)L'/c. \qquad (14)$$

where *L'* is taken to be invariant under rotation. The prediction (14) is obviously not consistent with the null result. It is usually now argued[17] that the implicit MM claim that each arm has the same length relative to *S'* must be abandoned, and replaced with the statement that, both before and after rotation, the length of the arm pointing in the direction of the ether wind must be shortened to $\gamma^{-1}L'$. (And indeed, a recalculation of (14) under this constraint yields $\delta' = (\delta^{rot})' = 0$, although again it is not the only such constraint on the rest dimensions of the interferometer which results in the vanishing of the delay times.) Thus, it is effectively being claimed *that a combination of length contraction and the Galilean rule for transforming velocities is compatible with the null MM result*. This is incorrect, and very misleading. What the argument leading to (13) shows is simply that the null result of the MM experiment is inconsistent with the Galilean law of composition of velocities (as will become clearer in the next section)[18]. To believe otherwise is both to deny the validity of the relativity principle and to uphold a wholly untenable picture of length contraction.[19]

So much for the explanation of the null result. Before we turn to the historical responses of FitzGerald and Lorentz, it behoves us to briefly review the kinematical implications of the MM experiment from a modern point of view.

4. Michelson-Morley kinematics.

Given the isotropy of the 'relative 3-space' in $S$, we can write the linear (standard configuration) coordinate transformations between $S$ and $S'$ quite generally as:

$$x' = C_L^{-1}.(x - vt) \qquad (15a)$$
$$y' = C_T^{-1}.y \qquad (15b)$$
$$z' = C_T^{-1}.z \qquad (15c)$$
$$t' = [D.(1 - \alpha v)]^{-1}.(t - \alpha x) \qquad (15d)$$

where again $C_L$, $C_T$ are the longitudinal and transverse 'deformation' factors, respectively, and $D$ is the time dilation factor. The factor $\alpha$ is the 'relativity of simultaneity' factor: it determines the degree to which the simultaneity relation in $S'$ differs from that in $S$. (Notice that $t' = 0$ implies $t = \alpha x$.) This factor is determined *solely* by the conventions introduced for synchronising distant clocks in both the frames in question.

The general transformations (15a,b,d) give rise to the following velocity transformation rules:

$$u_x' = \frac{D.(1 - \alpha v)(u_x - v)}{C_L.(1 - \alpha u_x)} \qquad (16a)$$

$$u_y' = \frac{D.(1 - \alpha v)u_y}{C_T.(1 - \alpha u_x)} \qquad (16b)$$

where $u_x$ is the $x$-component of the velocity of some body as measured relative to $S$, and $u_x'$ is the $x'$-component of the velocity of the same body relative to $S'$, etc.

Now the null result of the MM experiment demonstrates that the two-way speed of light is isotropic relative to the laboratory frame $S'$.[20] But it does not establish that this speed is numerically equal to $c$, the two-way light-speed in $S$. From the linearity of the transformations (15) it follows from the constancy of the light-speed in $S$ that the two-way light-speed in $S'$ is also independent of the speed of the source. Given that it is also isotropic, it is natural to adopt in $S'$, just as in $S$, the Einstein convention for synchronising clocks, which renders the one-way light-speed isotropic in $S'$.

Let us denote by $c'$ the two-way light-speed in $S'$. Assuming Einstein synchrony in $S'$, and hence using (16a) with $u_x = c$   $u_x' = c'$, $u_x = -c$   $u_x' = -c'$, one obtains

$$\alpha = v/c^2 \qquad (17)$$
$$D = c'^2 C_L/c. \qquad (18)$$

The relationship between $C_L$, $C_T$, is fixed by (12), so combining this result with (15), (17) and (18), we arrive at the following form of the coordinate transformations consistent with the MM null result:

$$x' = k\gamma.(x - vt) \qquad (19a)$$
$$y' = ky \qquad (19b)$$
$$z' = kz \qquad (19c)$$
$$t' = (k\gamma c/c').(t - vx/c^2) \qquad (19d)$$

where $|v| < c$. The temporal transformation (19d) can be rewritten as

$$t' = (^2/D)(t - vx/c^2) \qquad (20)$$

in which form it is clearer that the MM result does not constrain the dilation factor[21]. The 'MM-related' transformations (19) were derived in a somewhat different manner by Brown and Maia in 1993, and are consistent with the findings of Robertson in the section of his classic 1949 analysis of experimental kinematics dealing with the MM experiment.[22] Several points concerning their significance need to be stressed.

(i) The value of the dimensionless factor $k = k(v)$ in (19) is not fixed by convention, but experiment: it will affect in particular the measurable degree to which rods deform in motion. We have from (19a, b) that

$$C_L = (k)^{-1} \qquad (21)$$
$$C_T = k^{-1}. \qquad (22)$$

We shall have more to say later about the experimental significance of $k$; in the meantime we note that unless $k$ is an even function of $v$, the deformation effects will be anisotropic relative to $S$.

(ii) In the derivation of (19) on the basis of the MM result, no appeal is made to the relativity principle. Indeed, in general the deformation effects, say, defined for motion relative to *arbitrary* frames may differ considerably from those defined for motion relative to $S$ and exhibited in (21) and (22). For example, given the choice $k =$ , it follows from (21) that $C_L =$ $^{-2}$ defined relative to $S$, but inverting (19) shows that $C_L = 1$ relative to $S'$ when the 'moving' rod is at rest relative to $S$, even when $c' = c$.

(iii) The transformations (19) are in general incompatible with the principle of 'reciprocity', i.e. the requirement that when frame $S'$ moves with speed $v$ relative to $S$, then $S$ moves at speed $–v$ relative to $S'$. It is easy to show that reciprocity holds for (19) only if $D = $ , or equivalently $c = c'$. But reciprocity (once one fixes the synchrony convention in both frames) is an empirical issue, not one of logical necessity, and the fact that it holds in the non-relativistic approximation does not imply that it holds exactly.

(iv) It was noted in the previous section that the Galilean rule of transformation of velocities leads to (14), which is inconsistent with the null MM result. The reader will also see from (16a,b) that even if we choose the synchrony convention in $S'$ in such a fashion that $\alpha = 0$ (which we must do if we want Galilean kinematics to emerge, and which does not affect the contraction factors defined relative to $S$) the MM result $C_L$ $C_T$ rules out consistency with the Galilean rule.

(v) It was shown by Poincaré in 1905[23] that the Maxwell field equations for the vacuum are *exactly* form invariant (covariant) under the transformations obtained from (19) when $c = c'$, i.e. under the familiar Lorentz transformations with the multiplicative factor $k$. (Naturally, in order to obtain this result, changes must be made to the now standard relativistic transformations for the charge density $\rho$, and the components of the current *j* , as well as the electric and magnetic field strengths *E*, *B*.[24]) We shall refer to these as the *k-Lorentz transformations*. By 1909 it was known, through the work of E. Cunningham and H. Bateman, that the field equations are actually covariant under a 15-parameter group of 'conformal' transformations, which include the *k*-Lorentz transformations (as well as certain nonlinear ones).[25] Now it follows from what might be called Poincaré's 'limited conformal invariance' result associated with the *k*-Lorentz transformations, that *null* outcomes must be obtained in *all conceivable* ether wind experiments performed in an earth-bound laboratory, involving optical effects or more generally electromagnetic effects that can be accounted for solely by the Maxwell field equations. At least this is the case if all rods and clocks deform and dilate in conformity with the quantities in (15) determined by the *k*-Lorentz transformations.

## 5. The FitzGerald deformation hypothesis

In 1889, the American journal *Science* published in its May 2$^{nd}$ issue a brief letter by the Irish physicist George Francis FitzGerald[26], entitled "The Ether and the Earth's Atmosphere"[27]. (FitzGerald had chosen this journal in an attempt to capture the attention of Michelson and Morley, and also because he had at the time fallen out with the Royal Dublin Society, whose proceedings were his usual vehicle of publication.) It contained a sensational suggestion concerning "almost the only hypothesis" that could reconcile the 1887 MM experiment with the earlier, first-order, ether wind experiments. Here is the first part of the letter.

> I have read with much interest Messrs Michelson and Morley's wonderfully delicate experiment attempting to decide the important question as to how far the ether is carried along by the Earth. Their result seems opposed to other experiments showing that the ether in the air can be carried along only to an inappreciable extent. I would suggest that almost the only hypothesis that could reconcile this opposition is that the lengths of material bodies changes, according as they are moving through the ether or across it, by an amount depending on the square of the ratio of their velocities to that of light.

FitzGerald's *Science* letter was to remain virtually unknown until the historian Steven G. Brush drew attention to it in 1967[28]—not even its author was sure it had appeared in print! (Joseph Larmor, a friend and colleague of FitzGerald who was responsible for editing his collected works in 1902[29], was unaware of its existence[30].) Rather than publish the hypothesis in a more prominent journal, FitzGerald sought to promote it primarily by way of lectures and private communications with colleagues. As it happens, he had first voiced the hypothesis in 1889 during a visit to the Liverpool home of Oliver Lodge[31], and the first references to it in print were in papers Lodge published on optics in 1892 and 1893[32]. In the first, Lodge takes FitzGerald to be supposing "the size of bodies to be a function of their velocity through the ether". In the second paper, to which we return later, FitzGerald's claim is taken to be that the size of bodies "may be a function of their direction of motion through the ether; and accordingly that the length and breadth of Michelson's stone supporting block [on which the interferometer is mounted] were differently affected ...".

In a significant 1988 historical treatment of FitzGerald's hypothesis, Bruce J. Hunt claimed that "there is no reason to think that the idea that flashed on him in Lodge's study involved anything other than a simple [longitudinal] contraction"[33]. Although we may never know with certitude exactly what FitzGerald had in mind in 1889, Hunt's claim is surely questionable. It is not implausible that both FitzGerald and Lodge were aware from the beginning (as Lodge clearly was later) of the fact that the MM experiment does not require exclusively longitudinal contraction, and can be accounted for by a more general motion-induced deformation. Such awareness would certainly go a long way towards explaining both FitzGerald's and Lodge's initial formulations of the hypothesis. The "vagueness" of these formulations had been noted in 1966, with apparent puzzlement, by Alfred M. Bork, who wrote that they "do not state just what contraction is involved, in terms of mathematical details". Bork indeed warned that Lodge's 1893 paper "might easily be interpreted to indicate the effects in both length and breadth" of the moving body, which was exactly how FitzGerald's 1889 note was read by Mamone Capria and Pambianco in 1992, to their credit.[34] It is noteworthy that FitzGerald himself never (not even in his correspondence with Lorentz and Larmor) used the words "contraction" or "shortening", but referred to the length of the body changing depending on the orientation of the body relative to the direction of motion through the ether, with a consequent alteration in the size of the body.

In his last two reminiscences concerning the 1889 discussion with FitzGerald at his Liverpool home[35], Lodge recalls FitzGerald accepting his suggestion that the effect of motion on Michelson's stone slab might be a shear distortion. Bork, who in 1966 found the discrepancy between this suggestion and the familiar contraction hypothesis "astounding", thought it "perhaps more likely … [that Lodge] is projecting a view of his own"[36]. What is clear is that by his own admission[37], Lodge found it difficult to believe in a distortion (including pure longitudinal contraction) that did not preserve volume; by 1913 he openly defended the case of shear-distortion (corresponding to the choice $k = -1/3$ from our previous kinematical discussion). But Lodge's several accounts, from 1892 to 1931, of FitzGerald's *own* position in 1889 are remarkably consistent, and in our opinion deserve the benefit of the doubt. They do not attribute shear distortion to FitzGerald other than as one possible kind of deformation amongst others. It is surely worth noting in this connection that there appears to be no evidence of FitzGerald complaining that his hypothesis was misconstrued by either Lodge or the Dutch physicist H.A. Lorentz (see below), whose similar renderings of his hypothesis may well have become standard[38]. Nonetheless, a full analysis of the plausibility of Hunt's 1988 claim must take into account FitzGerald's view as to the underlying physical cause of the deformation, to which we turn in section 8 below.

By the time FitzGerald died in 1901, just under 50 years of age, the length-change hypothesis had gained considerable circulation, and his name, along with that of Lorentz, was usually associated with it. But in the mid-nineties, FitzGerald did not find much sympathy for the idea north of the Channel. He wrote to Lorentz in 1894 saying that he (FitzGerald) had been "rather laughed at for my view over here", and that Lodge had only mentioned the deformation hypothesis in his papers of 1892 and 1893 as a result of "reiterated positiveness" on FitzGerald's part.[39] (In his 1909 account of the 1889 meeting with FitzGerald, Lodge stated that the latter's deformation suggestion "bore the impress of truth from the start"[40]. But it is fairly clear in his 1892 and 1893 papers, particularly the latter, that Lodge was initially far from convinced of its validity.[41]) When, also in 1894, FitzGerald mentioned his hypothesis to R.T. Glazebrook and J.J. Thomson during a stay in Cambridge, his listeners found it—in Glazebrook's subsequent and regretful words—"the brilliant baseless guess of an Irish genius".[42]

6. Lorentz's deformation hypothesis

Unaware of both FitzGerald's 1889 Science letter and Lodge's 1892 paper, Hendrik Antoon Lorentz had independently[43] hit on essentially the same idea in 1892. Late in that year, after providing a detailed analysis of the effect of motion on the electrostatic forces maintaining a collection of charged particles in equilibrium, Lorentz wrote a paper entitled "The Relative Motion of the Earth and the Ether"[44]. Here, he introduced a second-order contraction effect to explain the null MM result, after years of puzzling over its significance. Yet another two years were to pass before Lorentz discovered FitzGerald's priority when reading the second of Lodge's papers alluded to above.

Lorentz's 1892 paper initially gives for the pre-rotation delay times in arms $A$ and $B$ of the Michelson interferometer the expressions (in our notation)

$$T_A \sim (2L/c).(1 + v^2/c^2) \qquad (23)$$
$$T_B \sim (2L/c).(1 + v^2/2c^2). \qquad (24)$$

Lorentz attributes these results to Maxwell. Although expressions (23), (24) are the second-order versions of our equations (1), (2) for $L_A = L_B = L$, it is initially ambiguous in the context of Lorentz's thinking as to which of the two frames $S, S'$ the times and lengths in (23) and (24) refer. Such measures were of course taken to be frame-independent, giving rise to a predicted delay

$$= T_A - T_B \sim Lv^2/c^3 \qquad (25)$$

which is proportional, as we have seen, to the shift in the interference pattern originally expected under a 90° rotation of the apparatus.

In order now to account for the null result of the MM experiment, Lorentz replaces $L$ in (25) by $(1 - \alpha)L$. It is now clear that $L$ in (23), (24) is the rest length of both arms $A$ and $B$ relative to the ether frame $S$, and that $\alpha = \alpha(v)$ is the decrease in the longitudinal length, defined relative to $S$, of a rod of unit rest length, produced by the motion $v$. (We note again that if such a length change affects all the rigid matter in the laboratory as well as the arms of the interferometer, as Lorentz clearly believed, then we can equate the rest length $L$ above with $L'$, i.e. the rest length of the arms relative to the laboratory frame $S'$, as in equation (11) above.) We must regard then the expressions (25), (26) and (27) as referring in the first instance to spatiotemporal measures defined relative to the privileged ether frame $S$. Ignoring terms of the order $v/c^2$, Lorentz noted that the delay effectively vanishes (and "with it the whole difficulty") if $\alpha = v^2/2c^2$. Observe that again the entire argument need make no appeal to the Galilean law of transformation of ray velocites: all that is involved is the constancy of the light-speed relative to $S$ and the possibility of length contraction. Nor did Lorentz explicitly evoke the Galilean law. Nonetheless, however appealing the possibility is that he avoided it—and the attendant difficulties outlined in sections 3 and 4 above—in his implicit thinking, the issue cannot be regarded as settled.[45]

We can at any rate now write Lorentz's solution in terms of our previous notation: $C_L \sim (1 - v^2/2c^2)$, $C_T \sim 1$. Yet Lorentz was aware in 1892 that other deformation effects are also consistent with the MM experiment; in particular he mentioned that a purely transverse increase in the dimensions of bodies "would answer the purpose equally well". It was later in his 1895 book[46] on electromagnetic and optical phenomena associated with moving bodies (often referred to simply as the '*Versuch*') that the point was brought out most systematically. Here, Lorentz replaces the 1892 contraction factor by a combination of the longitudinal factor $C_L = (1 + \delta)$ and the transverse factor $C_T = (1 + \varepsilon)$. He claims now that the vanishing of the delay requires

$$\varepsilon - \delta \sim v^2/2c^2 \qquad (26)$$

which is consistent with our condition (13). Lorentz stresses that "the value of one of the quantities $\delta$ and $\varepsilon$ would remain undetermined". As well as the 1892 values $\delta = -v^2/2c^2$, $\varepsilon = 0$, he explicity refers to the possibilities $\delta = 0$, $\varepsilon = v^2/2c^2$ (transverse expansion), and $\delta = -v^2/4c^2$, $\varepsilon = v^2/4c^2$ (longitudinal contraction and transverse expansion), though clearly these are not the only possibilities.

From the 1895 *Versuch* to his 1906 New York lectures (published as *The Theory of Electrons* in 1909[47]), Lorentz would repeatedly refer to his hypothesis, introduced to account for the MM null result, as one concerning the "changes of dimensions" of a moving solid body, and not specifically the longitudinal contraction thereof. (He explicitly attributed the same hypothesis to FitzGerald in 1895 —after reading Lodge's 1893 paper—, as well as in 1899 and 1906.[48]) The number of commentators who have appreciated the difference between length contraction and Lorentz's hypothesis is remarkably small, the greatest emphasis in the case of Lorentz's 1892 treatment having been made on the 1980s by Nancy J. Nersessian.[49]

7. FitzGerald on molecular forces in moving bodies

Both FitzGerald and Lorentz were clearly aware that the deformation hypothesis required some degree of theoretical underpinning if it were not to be dismissed as blatant trickery, or at least entirely *ad hoc*. Independently, they appealed to the possible effects of motion (relative to the ether) on the forces holding the molecules of rigid bodies in equilibrium, in analogy with the corresponding effect on 'electric' forces. Here is what FitzGerald said in his 1889 *Science* letter:

> We know that electric forces are affected by the motion of the electrified bodies relative to the ether and it seems a not improbable supposition that the molecular forces are affected by the motion and that the size of the body alters consequently.

An important insight into what FitzGerald had in mind was provided in the 1988 historical study due to Hunt mentioned earlier[50]. Hunt pointed out that some months before he wrote his 1889 letter to *Science*, FitzGerald had been in communication with Oliver Heaviside. The correspondence concerned Heaviside's important 1888 analysis, based on Maxwell's equations, of the distortion, or anisotropy, of the electromagnetic field of a charged spherical body due to motion relative to the ether—leading a few years later to the discovery of the 'Heaviside ellipsoid' by G.F.C. Searle[51]. In a letter to Heaviside written in January 1889, FitzGerald made the remarkable suggestion that the distortion result might be applied "to a theory of the forces between molecules". (The letter contains another remarkable suggestion, *viz*. that the speed of light might be a limiting speed. Was this the first time this notion was advanced?[52])

The only question is, how? Hunt took FitzGerald's suggestion as an indication that he thought that intermolecular forces might be electromagnetic. But even though, as Hunt stresses, FitzGerald regarded all physical forces as having their origin in disturbances of a single ether, it is not clear that Hunt's extrapolation is correct; no explicit statement to this effect seems ever to have been made by FitzGerald.[53] (It has also been stressed by Andrew Warwick that such a statement is hard to reconcile with Lodge's 1893 account of FitzGerald's hypothesis[54].)

The formulas Heaviside (and later J.J. Thomson) derived for the electric and magnetic field strengths surrounding a uniformly moving charged body are precisely those obtained more simply by assuming the validity of Maxwell's equations relative to the rest frame of the charge and using the known transformation properties of the fields—i.e. by doing it the modern way.[55] The formula for the electric field is:

$$\boldsymbol{E} = (q\boldsymbol{r}/r^3)(1 - v^2/c^2)(1 - v^2 \sin^2/c^2)^{-3/2} \qquad (27)$$

where $\boldsymbol{E}$ is evaluated at a point with displacement $\boldsymbol{r}$ from the centre of the charged body and is the angle between $\boldsymbol{r}$ and the direction of motion. Hunt interpreted the Heaviside formula as directly demonstrating a longitudinal "contraction of the electrostatic field"[56], but again it is not clear that this is how it was understood in 1888-9, by FitzGerald or anyone else. Hunt himself stressed a complicating factor for his thesis: that the clarification of the Heaviside ellipsoid was made in letters to Heaviside by Searle only in 1892, and then only published in 1896[57]. It might also be worth noting that whereas today the Heaviside ellipsoid is usually taken to refer to the surfaces in the field corresponding to constant values of the coordinate-independent quantity $(E^2 - B^2)$[58], for Searle it referred to the shape of the charged body in motion (assuming it not to be a point particle, and that the charge is in an equilibrium distribution on the surface of the body)[59].

Hunt claimed that by "combining Heaviside's formula with what he had already believed about the electromagnetic nature of intermolecular forces, FitzGerald would have been able to predict the contraction effect had he never heard of the Michelson-Morley experiment"[60]. In our opinion, such a claim should be viewed with a degree of skepticism on a number of counts. However, the merit of Hunt's 1988 study cannot be denied. It is to have shown that through his knowledge of Heaviside's 1888 work on the distortion of the electromagnetic field associated with moving charges, FitzGerald could advance the "not improbable" idea of intermolecular forces being also affected in some way by the ether wind[61]. It is plausible that he actually hit on his solution to the MM conundrum as a result of awareness of Heaviside's work. He may well have done so before the celebrated discussion in Lodge's study in March or April of 1889.

8. Lorentz on molecular forces

Both in 1892 and 1895 it was independently argued by Lorentz that it is "not far-fetched" to infer an ether-wind effect on molecular forces given a related effect on electrostatic forces. But his argument was different from FitzGerald's in two respects.

First, Lorentz employed the analogy not with the Heaviside result concerning a moving charged body (which he may not have known), but a quantitative treatment he himself had given earlier in 1892 of the motion-induced deformation of a collection of charged particles held in equilibrium. This treatment did not involve solving Maxwell's equations in the way that Heaviside's 1888 work did, and was considerably less direct, but a fuller explanation of the distinction must be deferred to a subsequent paper.

The second difference concerned the very nature of the analogy between electrostatic and intermolecular forces. It is true that, like FitzGerald, Lorentz entertained the idea that "molecular forces are also transmitted through the ether", the way Lorentz put it in 1895. Let us call this the *weak* molecular forces hypothesis. But in both 1892 and 1895 Lorentz also toyed with a *strong* version of the hypothesis.[62] (Note that in the weak version, it may well be asked whether the speed of propagation of intermolecular interactions is expected also to be $c$, relative to the ether rest frame. Under the *mechanical* interpretation of the ether as a ponderable elastic body, it is difficult to imagine otherwise, but it is worth noting that of this interpretation Lorentz himself was no advocate.[63])

The strong version claimed that the manner in which the molecular forces are affected by motion is *exactly* that associated with electrostatic forces. It goes without saying that if the molecular force is ultimately electromagnetic in nature, then the strong claim is trivially valid. But unlike Larmor in 1895 and 1900, for instance, the ever-cautious Lorentz never appears to have adopted this reduction thesis, and we have already questioned whether FitzGerald did. Nor was Lorentz's commitment to the strong version of the molecular forces hypothesis seemingly on a par with that concerning the weak version. There was, as he expressly put it in 1895, "no reason" for adopting the former. But his point was merely that *if* the strong version is true, then a deformation effect consistent with his mysterious formula (28) automatically ensues.

In a letter to Einstein written in 1915 and unearthed many years later by A.J. Kox[64], Lorentz admitted that he had arrived at the deformation hypothesis shortly before he developed the plausibility argument based on molecular forces. But he expressed regret that he had not emphasised the dynamical argument more from the beginning: had he done so "the hypothesis would have made less an impression of having been devised *ad hoc*."

Returning finally to the early years, it seems to have been Lorentz's belief both in 1892 and 1895 that a collection of charged particles held in equilibrium under electrostatic forces (insofar as this makes sense) undergoes a solely longitudinal contraction, according to the (relativistic) factors $C_L = ^{-1}, C_T = 1$. The validity of the strong version of the molecular forces hypothesis would therefore yield (to second order) the values $= -v^2/2c^2, = 0$. But Lorentz had already warned in 1892[65]:

> One may not of course attach much importance to this result; the application to molecular forces of what has found to hold for electric forces is too venturesome for that.

Furthermore, by 1899, it had become apparent to him that the strong version of the electrostatic analogy —or rather his construal of it—would after all allow for the more general deformation factors $C_L = (k\ )^{-1}, C_T = k^{-1}$, with $k$ unknown, consistent (to second order) with his relation (28) above. Lorentz wrote in 1899[66]:

> We need hardly remark that for the real transformation produced by translatory motion, the factor [$k$] should have a definite value. I see however no means to determine it.

## 9. Final remarks

Lorentz would attempt to show systematically in 1904 that the value of the scale factor $k$ is unity, and consequently that the effect of motion on rigid bodies is a purely longitudinal contraction. But the argument was convoluted and later not even Lorentz seemed to find it entirely convincing. As J. H. Jeans remarked some years later[67], the edifice Lorentz had created in analysing the electrodynamics of systems in motion "needed for its consolidation a theory which ultimately came from other hands". Those hands belonged to Henri Poincaré and, more significantly, Albert Einstein. But it was not just an appreciation of the true nature of time dilation and relativity of simultaneity that was needed, as well as of the otiose nature of the electromagnetic ether. It was also necessary to determine, in a transparent way, *what kind of deformation* is suffered by rigid bodies in motion.

*Acknowledgements*
My greatest debt is to Ronald Anderson S.J., who provided, besides regular encouragement, details both of recent historical studies related to this topic and of relevant letters written by FitzGerald. Arthur Miller also provided useful bibliographical details and historical clarifications during several discussions, and Michel Janssen kindly provided me with a copy of his remarkable 1995 Ph.D. thesis on Lorentz. Useful discussions of the pre-relativistic treatment of the MM experiment took place with Peter Torok and particularly Marco Mamone Capria. Comments on earlier versions of the paper were gratefully received from Tim Budden, Nancy Nersessian, Simon Saunders, Roland Sypel and Andrew Warwick. Thanks also go to undergraduate Oxford classes, over several years, in the Physics and Philosophy course, and particularly Marcus Bremmer, James Orwell, and Katrina Alexandraki for helping the author to pose some of the questions that gave rise to this study. Finally, support from the British Academy and the Leverhulme Trust is gratefully acknowledged.

---

[1] Albert Einstein, "Zur Electrodynamik bewegter Körper", Annalen der Physik 17, 891-921 (1905). A recent English translation can be found in John Stachel (ed.), *Einstein's Miraculous Year; Five Papers that Changed the Face of Physics* (Princeton University Press, 1998), pp. 123-160.

[2] For a clear discussion of these speeds and their distinct (non-relativistic) transformation properties, see C. Møller, *The Theory of Relativity*, (Oxford University Press, Delhi, 1972), section 1.6.

[3] See Albert A Michelson, "The Relative Motion of Earth and Ether", *Am J Sci*, 3rd ser., **22**, 120-129 (1881) and Albert A Michelson and Edward W Morley, "On the Relative Motion of the Earth and the Luminiferous Ether", *American Journal of Science,* 3rd ser., **34**, 333-345 (1887). For discussion of the historical background to these experiments, see R. S. Shankland, "Michelson-Morley experiment", Am. J. Phys. **32**, 16-35 (1964) and particularly L. S. Swenson, *The Ethereal Ether; A History of the Michelson-Morley-Miller Aether-Drift Experiments 1880-1930* (University of Texas Press, Austin, 1972). A more recent account of the 1881 experiment is given in Barbara Haubold, Hans Joachim Haubold and Lewis Pyensen, "Michelson's first ether-drift experiment in Berlin and Potsdam", in Stanley Goldberg and Roger H. Stuewer (eds.), *The Michelson Era in American Science 1870-1930* (American Institute of Physics, New York, 1988) pp. 42-54. A modern version of the MM experiment using a microwave cavity is found in A. Brillet and J.L. Hall, "Improved laser test of the isotropy of space", Phys. Rev. Lett. **42**, 549 (1979). (Thanks go to Claus Lämmerzahl for this reference.)

[4] A recent study of the history of the Fresnel drag coefficient is found in K. M. Pedersen, "Water-Filled Telescopes and the Pre-History of Fresnel's Ether Dragging", Arch. Hist. Exact. Sci. **54**, 499-564 (2000). An excellent treatment of the role of the coefficient in explaining the null first-order ether wind results is given by Michel Janssen and John Stachel, "The Optics and Electrodynamics of Moving Bodies", to appear in *History of 19th Century Physics*, Jed Buchwald (ed.), Istituto della Enciclopedia Italiana. See also Andrew Warwick, "On the role of the FitzGerald-Lorentz contraction hypothesis in the development of Joseph Larmor's electronic theory of matter", Archive for the History of the Exact Sciences **43**, 29-91 (1991).

[5] Derivations of the equations (1), (2) for quantities referred, as in our account, to the ether rest frame, are found for example in Arnold Sommerfeld, *Optics, Lectures on Theoretical Physics, Vol. IV* (Academic Press, New York, 1954), section 14, and Stanley Goldberg, *Understanding Relativity. Origin and Impact of a Scientific Revolution* (Clarendon Press, Oxford, 1984) pp. 450-451.

[6] Feynman's treatment of the MM experiment gives the impression that the purpose of the rotation is to render inconsequential the fact that arms *A* and *B* cannot have exactly the same length. (See R. P. Feynman, R. B. Leighton and M. Sands, *The Feynman Lectures on Physics, Vol I* (Addison-Wesley Publishing Co., Reading Mass, 1963), section 15-3.) Now it is true that the difference in the delay times given by (3) and (4) will be largely unaffected whether the arm lengths are exactly the same or very slightly different (see Sommerfeld *op. cit.* note 5 above, p. 78, footnote 1). But the more significant reason for the rotation is that it is only the fringe *shift* that provides in principle any evidence of the ether wind, as hopefully will become apparent.

[7] V. S. Soni, "Michelson-Morley analysis", Am. J. Phys. **57**, 1149-1150 (1989).

[8] But see in this connection the plausible story in M. Mamone Capria and F. Pambianco, "On the Michelson-Morley Experiment", Found. Phys. **24**, 885-899 (1994), footnote 6. Note that this was the second erroneous expression for $T_B$ provided by Michelson; in his 1881 (see note 2 above) he gave $T_B = 2L_B/c$.

[9] See the second term in (7') in Mamone Capria and Pambianco (*op. cit.* note 8), p. 891, which has been re-arranged.

[10] Equation (8) follows from equations (7') and (7") in *ibid*.

[11] These authors (*op. cit*. note 8) actually calculated, pp. 891-4, the exact expressions for the pre- and post-rotation delay times in the more general case where initially the arm *A* makes an arbitrary angle with the direction of the ether wind.

[12] See, e.g., Sommerfeld (*op. cit.* note 5), and Hendrik A. Lorentz, *Lectures on theoretical physics*. Vol 3 (MacMillan and Co., London, 1931), p. 204.

[13] See A. P. French, *Special Relativity* (Van Nostrand Reinhold UK, 1984), p. 54, for a clear indication of this.

[14] The claim, without proof, is found in Mamone Capria and Pambianco (*op. cit*. note 8) footnote 3. The basis of the claim was kindly explained by Marco Mamone Capria in private correspondence.

[15] Mamone Capria and Pambianco (*op. cit*., note 8, pp. 895-6) themselves give this argument in terms not of (12) but the usual relativistic length contraction, and in the case where $L_A' = L_B' = L'$. In fact these authors want to show that under these conditions the delay time itself vanishes. But it is easy to check in the case of the more general deformation associated with (12), that (i) the angle of tilt of the beam splitter is the same, and (ii) even when the lengths of the arms of the interferometer are not the same relative to the lab frame *S'*, the fringe shift (6)— understood in terms of (3) and (4)— vanishes, whatever the value of the time dilation factor.

[16] Robert Mills, *Space, time and quanta: an introduction to contemporary physics* (W.H. Freeman, New York, 1994).

[17] See, for example, Møller (*op. cit*. note 2), pp. 27; H. Muirhead, *The Special Theory of Relativity* (Macmillan, London 1973), pp. 10-11; French (*op. cit*. note 14) pp. 54-56, 63-64.

[18] It seems that the same conclusion is found in H Melchor, "Michelson's Ether-Drift Experiments and their Correct Equations", in Goldberg and Steuwer (*op. cit*. note 3), pp. 96-99.

[19] Since each arm of the interferometer changes, according to the argument, its *rest* length under rotation, it follows that there is a violation of spatial isotropy in the laboratory frame *S'*, contrary to the relativity principle taken in conjunction with the assumption of spatial isotropy relative to *S*. Furthermore, the contraction is selective: it cannot hold for rigid rulers attached to the stone slab in the experiment, since otherwise it would lead to strictly no discernible difference between the lengths of the two arms of the interferometer relative to the laboratory frame *S'*, contrary to

assumption. (It is noteworthy that Møller, *op. cit*, note 2, p. 28, recognises that the contraction relative to *S'* cannot be detected, after having used it on p. 27 to account for the null result!)

[20] It has recently been shown, however, that a very contrived model of electrodynamics involving anistropic light propogation could in principle be consistent with MM-type experiments, when the orientation dependence of the lengths of rigid bodies used in the experiments—itself a prediction of the model—is taken into account. See Claus Lämmerzahl and Mark P. Haugan, "On the interpretation of Michelson-Morley experiments", to appear in Phys. Lett. A (2001); e-print gr-qc/0103052.

[21] It will be seen in a subequent paper that a variant of the MM experiment proposed by Liénard in 1898 would constrain the dilation factor, as Lorentz realised in 1899. (See Michel Janssen, "A Comparison between Lorentz's Ether Theory and Special Relativity in the Light of the Experiments of Trouton and Noble." Ph.D. Thesis, University of Pittsburgh, 1995; pp. 187-8.)

[22] Harvey R. Brown and Adolfo Maia Jr., "Light-speed constancy versus light-speed invariance in the derivation of relativistic kinematics", Brit. J. Phil. Sci. **43**, 381-407 (1993); H. P. Robertson, "Postulate *versus* observation in the special theory of relativity", Rev. Mod. Phys. **21**, 378-82 (1949).

[23] Henri Poincaré, "Sur la dynamique d'électron", Rendiconti del Circolo Matematico di Palermo **21**, 129-175 (1906); the paper was submitted in July 1905. (Poincaré, like Lorrentz, denoted our *k* factor by *l*.)

[24] These changes involve a multiplicative factor $k^{-2}$ being introduced into the standard transformations for the components of the electric and magnetic field strengths, along with a factor $k^{-3}$ for the charge density and the components of the current.

[25] For a treatment of the conformal covariance of Maxwell's field equations, with good historical bibliography, see T. Fulton, F. Rorlich, and L. Witten, 'Conformal Invariance in Physics', Rev. Mod. Phys. **34** (3), 442-457 (1962). A less technically demanding treatment of the covariance result is found in G. Rosen, "Conformal Invariance of Maxwell's Equations", Am. J. Phys. **40**, 1023-1027 (1972).

[26] An excellent summary of FitzGerald's work and personality is found in Oliver Lodge, "G. F. FitzGerald", Proc. Roy. Soc. **75**, 152-160 (1905).

[27] George F. FitzGerald, "The Ether and the Earth's Atmosphere", Science **13**, 390 (1889). The letter is reprinted in Stephen G. Brush, "Note on the History of the FitzGerald-Lorentz Contraction", *ISIS* **58** 230-232 (1967) and the greater part of it in John S. Bell, "George Francis FitzGerald", 1989 lecture, abridged by Denis Weaire in Physics World, September 1992, 31-35. For the historical background to this letter, we have relied heavily on Alfred M. Bork, "The 'FitzGerald' Contraction", *ISIS*, **57**, 199-207 (1966), Bruce J. Hunt, "The Origins of the FitzGerald Contraction", Brit. J. Hist. Sci. **21**, 67-76 (1988) and Warwick (*op. cit.* note 4).

[28] See Brush, *op. cit*, note 27.

[29] *Scientific Writings of the Late George Francis FitzGerald*, Joseph Larmor (ed.) (Hodges, Figgis; Dublin, 1902).

[30] See Hunt (*op. cit.* note 27).

[31] Lodge was to mention this episode at least four times in print: Oliver Lodge, (*op. cit.* note 26); *The Ether of Space* (Harper and Brothers, London, 1909), p. 65; "Continuity" (Presidential address), British Association Report (1913), 3-42 (see pp. 25-26); *Past Years; an autobiography* (Hodder and Stoughton, London, 1931), pp. 204-205. In the last two cases, Lodge provided slightly different verbatim accounts of the discussion.

[32] Oliver Lodge, "On the present state of our knowledge of the connection between ether and matter: an historical summary", Nature (16 June 1892) **46**, 164-165; "Aberration problems", Phil. Trans. **184A**, 727-804 (1893); see in particular pp. 749-750.

[33] Hunt (*op. cit.* note 27).

[34] See Bork (*op. cit.* note 27), and Mamone Capria and Pambianco (*op. cit.* note 8), p. 895. Apart from these latter authors, it is very difficult to find modern commentators who attribute a deformation, rather than contraction, hypothesis to FitzGerald. However, as we shall see, this is much less the case for the decades following 1890.

[35] See Lodge (1913, *op. cit.* note 31) and (1931 *op. cit.* note 31).

[36] Bork (*op. cit.* note 27), p. 206.

[37] Lodge (1931 op. cit. note 31)

[38] Hicks wrote in 1902: "Amongst the various explanations advanced to account for the supposed [MM] null result …, the best known and accepted is that first proposed, I believe, by G. F. FitzGerald, viz., that the very motion of a solid through the aether

produces a small extension perpendicular to the direction of drift, or contraction in the direction …". (W. M. Hicks, "On the Michelson-Morley Experiment relating to the Drift of the Æther", Philosophical Magazine and Journal of Science, **III**, Sixth Series, 9-42 (1902), pp. 38-9. It is noteworthy that according to Hicks' own analysis of the MM experiment (*ibid*) the FL deformation could not account for the null result.) In his famous 1921 encyclopedia article on relativity theory, Pauli likewise described FitzGerald's hypothesis in terms of a shape deformation, in fact of the quantitative kind proposed by Lorentz (see below). (Wolfgang Pauli, *Theory of Relativity* (Dover, New York, 1981), pp. 2, 3.)

[39] See Bork (*op. cit*. note 27) and Bell (*op. cit*., note 27).

[40] Lodge (1909 *op. cit*. note 31), p. 65.

[41] In this connection see also Bork (*op. cit*. note 27), p. 200.

[42] R. T. Glazebrook, contribution to the obituary "H. A. Lorentz", Nature (25 February 1928) **121**, 287-288.

[43] The account of Lorentz's relevant work found in Whittaker's famous 1910 history of ether theories erroneously claims that Lorentz adopted FitzGerald's hypothesis (that "dimensions of material bodies are slightly altered when they are in motion relative to the aether") some months after its appearance in Lodge's 1892 paper (*op. cit*. note 32). See E.T. Whittaker*, A History of the Theories of Aether and Electricity*. Vol. I (Longmans, Green: London. 1910). Bork (*op. cit*. note 27), pp. 432-433, appears to regard the error as a deliberate distortion. We note that it is repeated in Whittaker's contribution to the 1928 Nature obituary for Lorentz (see note 42).

[44] H.A. Lorentz, "De relatieve beweging van de aarde en den aether." Koninklijke Akademie van Wetenschappen te Amsterdam. Wis- en Natuurkundige Afdeeling. Verslagen der Zittingen **1** (1892-93): 74-79. Reprinted in translation: "The Relative Motion of the Earth and the Ether", in: *Collected Papers* , P. Zeeman and A. D. Fokker, (eds.), Vol. 4 (The Hague: Nijhoff, 1937); pp. 219-223. This paper appeared in the 26 November, 1992 issue of the Amsterdam Academy proceedings; yet as late as August of the same year, Lorentz had written to Lord Rayleigh (see the quotation at the head of the present paper) admitting that he was "totally at a loss" to explain the contradiction between the MM result and Fresnel's hypothesis of a stagnant ether. See Kenneth F. Schaffner, *Nineteenth-Century Aether Theories* (Pergamon Press, Braunschweig, 1972), p. 103.

[45] Some historians have indeed speculated that Lorentz's 1892 explanation of the MM experiment *was* based on the non-relativistic addition law for velocities; see Arthur I. Miller, *Albert Einstein's Special Theory of Relativity* (Addison-Wesley, Reading, Mass, 1981), p. 31, and Nancy J. Nersessian, " 'Why wasn't Lorentz Einstein?' An Examination of the Scientific Method of H. A. Lorentz", Centaurus **29**, 205-242 (1986), p. 222.

[46] H.A. Lorentz, *Versuch einer Theorie der electrischen und optischen Erscheinungen in bewegten Körpern* (Brill, Leyden, 1895). An English translation of the Introduction is found in Schaffner (*op. cit.* note 44), pp. 247-254. A translation of sections 89-92 dealing with the MM experiment is found in H.A. Lorentz, A. Einstein, H. Minkowski and H. Weyl, *The Principle of Relativity* (Methuen & Co., London 1923), pp. 3-34.

[47] H.A. Lorentz, *The Theory of Electrons and its Applications to the Phenomena of Light and Radiant Heat* (B.G. Teubner, Leipzig, 1909), p. 195.

[48] "In order to explain the negative result of this experiment FITZGERALD and myself have supposed that, in consequence of the translation, the dimensions of the solid bodies serving to support the optical apparatus, are altered in a certain ratio." H. A. Lorentz, "Simplified Theory of Electrical and Optical Phenomena in Moving Systems", Koninklijke Akademie van Wetenschappen te Amsterdam, Section of Sciences, Proceedings 1 (1898-99): 427-442. (This 1899 paper was reprinted in Schaffner (*op. cit.* note 44), pp. 255-273 and was a slightly revised translation of an earlier paper: "Vereenvoudigde theorie der electrische en optische verschijnselen in lichamen die zich bewegen." Koninklijke Akademie van Wetenschappen te Amsterdam. Wis- en Natuurkundige Afdeeling. Verslagen van de Gewone Vergaderingen 7 (1898-99): 507-522.) See also Lorentz (1909 *op. cit.* note 47), p. 195, for similar remarks.

[49] At this point we cite those historical accounts known to us which recognise this difference. Early recognition is found in L. Silberstein, *The Theory of Relativity* (MacMillan, London, 1914), p. 78, and in Pauli (*op. cit.* note 38), pp. 2, 3. More recent cases are Kenneth T. Schaffner, "The Lorentz electron theory and relativity", Am. J. Phys. **37**, 498-513 (1969); Miller (*op. cit.* note 26), p.31 and footnote 21, p. 97; Mamone Capria and Pambianco (*op. cit.* note 8), Janssen (*op. cit.* note 21), section 3.2.2, and Janssen and Stachel (*op. cit.* note 3). The difference in the case of Lorentz

in 1892 was however emphasised most strongly by Nancy J. Nersessian, *op. cit.* note 45, and " 'Ad hoc' is not a four-letter word: H. A. Lorentz and the Michelson-Morley experiment", in Goldberg and Stuewer (1988, *op. cit.* note 3), pp. 71-77.

[50] Hunt (*op. cit.* note 27).

[51] Oliver Heaviside, "The electro-magnetic effects of a moving charge", Electrician **22**, 147-148 (1888); G.F.C. Searle, "Problems in electrical convection", Phil. Trans. **187A**, 675-713 (1896).

[52] See Hunt (*op. cit.* note 27) and Bell (*op. cit.* note 27).

[53] It was noted by Bell (*op. cit*, note 27) that in his 1889 Science letter, FitzGerald does not explicitly equate molecular forces with electromagnetic ones.

[54] Warwick (*op. cit.* note 4), p. 41.

[55] See for example W.G.V. Rosser, *Introductory Relativity* (Butterworth, London, 1967), p. 232. It is interesting that by the mid-1890's, Joseph Larmor was already capable of providing the 'modern' derivation: see Warwick (*op. cit.* note 4).

[56] Hunt *op. cit.* note 27.

[57] See Searle (*op. cit.* note 51).

[58] See for instance Møller (*op. cit.* note 2) p. 159.

[59] In speculating as to why FitzGerald did so little to publicise his solution to the MM enigma, Hunt speculated (*op. cit.* note 27, p. 75) that he "may also have been put off … by confusion about the directional variations of the forces between charges moving together, the intricacies of which were not cleared up until 1892 and were not well known for several years after that …". Hunt is presumably referring to the work of Searle. Even if Searle did provide the crucial clarification in this sense, it is difficult to reconcile Hunt's speculation with his own claim, mentioned in section 4 above, that FitzGerald deformation hypothesis involved, from the start, a strict longitudinal contraction.

[60] Hunt (*op. cit.* note 27).

[61] Nersessian (*op. cit.* 1988, note 49) was presumably unaware of these details when she wrote that there were "no *good reasons*" for FitzGerald's "contraction" hypothesis (p. 74), and that FitzGerald failed to pursue his speculation that the ether wind affected intermolecular forces "because he had no theory into which he could incorporate the speculation" ( fn. 26, p. 77).

[62] The distinction between the two versions of the molecular forces hypothesis was appreciated by Miller (*op. cit.* note 45) p. 30.

[63] Indeed, in Lorentz's view of the ether as composed of a new kind of imponderable material, it is no longer obvious that the light-speed should be constant in the ether rest frame, as of course Lorentz assumed. See S. Saunders and H.R.Brown, "Reflections on Ether", in S. Saunders and H.R. Brown (eds.) *The Philosophy of Vacuum* (Clarendon Press, Oxford 1991), pp. 27-63.

[64] See Janssen (*op. cit.* note 21, p. 197). The letter can be found in the Archief H.A. Lorentz, Rijksarchief Noord-Holland, Haarlem, The Netherlands; parts of it are quoted in Nancy J. Nercessian*, Faraday to Einstein: constructing meaning in scientific theories* (Martinus Nijhoff Publishers, Dordrecht, 1984), pp. 118-119, 172-173; in Nersessian (1986, *op. cit.* note 45, pp. 225, 232-233; and Nersessian (1988, *op. cit.* note 49), pp. 74-75.

[65] Lorentz *op. cit.* 1892, note 44.

[66] Lorentz *op. cit.* 1899, note 48.

[67] J.H. Jeans, contribution to the 1928 Nature obituary of Lorentz (note 42), pp. 287-288.